# AC loss in granular superconducting $MgB_2$ at radio frequencies

S. Sarangi,[a)] S. P. Chockalingam and S. V. Bhat

Department of Physics, Indian Institute of Science, Bangalore-560012, India

**Abstract:**

AC losses of granular superconducting $MgB_2$ were measured using non-resonant rf power absorption techniques. The presence of two band-gaps makes the temperature dependent ac loss pattern of $MgB_2$ different from other high $T_c$ superconductors like $YBa_2Cu_3O$ and BSCCO-2223. Josephson junction decoupling plays a major role in $MgB_2$ polycrystalline samples, both in the presence and absence of magnetic field. We studied the ac loss of $MgB_2$ samples with different crystalline properties, grain sizes and pressures and sintered under different physical and chemical conditions. The effects of frequency and amplitude of ac current on the ac loss are discussed in detail.

AC loss of a superconducting material is an important parameter in practical application because many applications involve an ac current or magnetic field. In type II superconductors, there are various loss mechanisms [1, 2, 3, 4, 5], namely, eddy-current losses, flux-flow losses, Josephson junction (JJ) decoupling losses, bulk-pinning hysteresis losses, surface pinning hysteresis losses, flux creep losses and flux-line cutting losses. The loss mechanism varies from sample to sample according to their microstructures. Recently we observed the ac losses in different granular superconductors in the rf range and concluded that the ac losses in granular superconductors are mainly due to Josephson junction decoupling [6].

There have been considerable amount of studies to understand how the ac losses in $MgB_2$ behave as a function of temperature and the frequency and amplitude of the applied ac field [7, 8, 9, 10]. But most of the studies are focused either in the low frequencies below 100 KHz or the frequencies in the microwave range. AC loss studies in the rf range and particularly in the granular superconductors are very rare in the literature. In this paper, we studied the ac loss of granular $MgB_2$ superconductor by studying the non-resonant power absorbed by the sample in the rf range [11, 6]. We found that the loss due to JJ decoupling is the main source of the ac losses in the granular $MgB_2$ sample. In this paper, we mainly depict our experimental study on the losses in $MgB_2$ polycrystalline samples in terms of the rf energy absorption phenomenon, demonstrating the influences of temperature, frequency, magnitude of rf field, grain size, pressure and sintering. The mechanisms of losses in polycrystalline $MgB_2$ are discussed and compared with other high $T_c$ superconductors like $YBa_2Cu_3O$ (YBCO) and BSCCO-2223 (BSCCO).

Commercially available $MgB_2$ powder (Alfa Aesar, 99.9% pure) was used for all the experiments. In order to investigate the effects of different particle sizes, a graded set of samples was prepared using a differential sedimentation method as described in [12]. The samples produced by this method had a graded range of particle sizes up to known maxima: less than 30 μm, less than 20 μm and less than 10 μm. We study the effect of pressure on the ac loss in granular $MgB_2$ sample by making pellets from the powder sample (less than 30 μm) at three different pressures: 10 MPa, 20 MPa and 30 MPa. To study the effect of sintering on the ac loss, we prepared two pellets made of polycrystalline samples of $MgB_2$. Both the pellets were prepared from the same phase of $MgB_2$ powder (grain size is less than 30 μm) with a pressure of 10 MPa. One of the two pellets was sintered at a temperature of 750 $^0C$ in flowing Argon for 5 hours to minimize the presence of Josephson junctions. The second pellet was not sintered to enhance the possibilities of Josephson junctions in it. The sizes of the pellets (cylindrical pellets of length: 7 mm and radius 3 mm) were kept constant for all the measurements. The applied rf frequency and the rf amplitude were set at 10 MHz and 700 mV for all the measurements.

Figure 1 shows the ac loss of granular $MgB_2$ sample. The superconducting transition is clearly visible in the figure at the temperature of 39 K. The result shows a sudden drop in the ac loss just below $T_c$. We interpret that the increase in the ac losses with lowering temperature is due to the increase of JJ critical current of the junctions present inside the granular sample. The inset in Fig.1 shows the ac loss of granular YBCO sample. AC loss in the case of YBCO shows one steady rise below $T_c$ and follows an Ambegaokar-Baratoff function for tunnel junction like behavior [13, 6], whereas the ac loss pattern of MgB$_2$ is fitted to the modified Ambegaokar-Baratoff function for tunnel junction suggested by Xu et al [14] and Clem [15] and adopted by Zhang et all [16], Darhmaoui et al [17] and Yan et al [18]. We have explained the ac loss behavior of YBCO granular sample in term of JJ decoupling in our recent paper [6]. The pronounced hump around 34 K is attributed with the two band-gaps in MgB$_2$ [19, 20]. Due to the presence of two band-gaps ($\Delta_\sigma$ and $\Delta_\pi$) in MgB$_2$, Josephson junction critical current among the superconducting grains varies as $J_\sigma$ and $J_\pi$ as shown in the figure. The ac loss from "A" to "B" is fitted to $J_\sigma$. The ac loss from "C" to "D" is fitted to ($J_\sigma+J_\pi$), as the ac loss in this region comes both from $J_\sigma$ and $J_\pi$. At the point "C", due to the rapid increment in $J_\pi$, rf energy (700 mV) becomes less than the Josephson junction decoupling energy. Josephson junctions with critical current $J_\pi$ do not break below the point "C", so the loss decreases from "C" to "B". Due to the presence of single band gap, YBCO shows a steady rise in the ac loss pattern. The ac loss in the case of YBCO sample is more than MgB$_2$ in the superconducting state. This proves that the Josephson junctions formed inside the YBCO pellet are stronger than the Josephson junctions formed inside the MgB$_2$ pellet.

Figure 2 shows a clear picture of the magnetic field dependent ac losses of MgB$_2$ polycrystalline samples at different temperatures below T$_c$. The phase of the magnetic field dependence ac loss goes through three separate and distinguished phase-reversals at temperatures of 27 K, 33 K and 37 K. The ac loss decreases with magnetic field in the range from −150 G to 150 G below 27 K, whereas the loss increases with magnetic field in between 27 K and 33 K. Above 33 K the ac loss decreases with magnetic field upto 37 K. In the transition region (above 37 K), the loss increases with magnetic field upto 39 K.

At 40 K the ac loss is independent of magnetic field. The application of a magnetic field and the variation of temperature alter the ac penetration depth of the sample, which in turn change the ac losses associated with it. This is the reason why the magnetic field dependent loss increases with field in the transition reason (above 37 K). At other temperatures (below 37 K), frequent JJ decoupling is the dominating factor. The Josephson junction critical current decreases with increasing magnetic field, so the decoupling energy ($E_j$) decreases with increasing magnetic field [6]. Due to the decreasing of $E_j$ with increasing magnetic field, ac loss at temperatures below 27 K decrease with increasing magnetic field. The magnetic field dependent ac loss at different temperatures from 27 K to 37 K can be explained according to the temperature dependent ac loss pattern from point "B" to point "D" in figure 1.

Figure 3 shows the details of the temperature dependent ac loss of granular $MgB_2$ at three different rf amplitudes. The ac loss increases drastically with increasing rf amplitude. In the case of 700 mV, the ac loss in the superconducting state is less than the normal state but in the case of 1300 mV, the ac loss in the superconducting state is more than the normal state. The rf current flow through the sample increases with rf voltage, increasing the current flow in the superconductor leads to the breaking of more number of Josephson junctions, so the ac loss increases with increasing rf amplitude and in the case of 1300 mV, due to the large number of Josephson junctions decoupling, the ac loss becomes more than the normal state.

Figure 4 shows the details of the temperature dependent ac loss of granular $MgB_2$ at three different grain sizes; <10 μm, <20 μm and <30 μm. The ac loss decreases with decreasing grain size. Inset shows the detail of the temperature dependent ac loss of granular $MgB_2$ at three different pressures; 10 MPa, 20 MPa and 30 MPa. The ac loss decreases with increasing pressure. Increasing pressure or decreasing grain size, both have the similar effect on the ac loss in granular samples. Increasing pressure or decreasing grain size decreases the number of weak links inside the pellets and makes some of the weak links stronger. In the first case, the loss decreases because of the total reduction in the number of Josephson junctions and in the second case the loss decreases because some of the weak links don't break with the same amplitude of applied rf current.

Figure 5 shows the ac losses of both the sintered and non-sintered $MgB_2$ pellets. Just below the transition temperature, sintered $MgB_2$ pellet shows the ac loss nearly zero down to the temperature of 4 K. The behaviors of the ac losses with frequency for both the samples are shown in the inset of Fig. 5 at a temperature of 10 K. The ac loss of the granular $MgB_2$ sample increases with frequency in the range of frequency from 5 MHz to 15 MHz with an average change of 770 X $10^{-15}$ *Watt* / Hz, but for the sintered sample, it increases with an average change of only 1 X $10^{-15}$ *Watt* / Hz. The ac loss in the sintered $MgB_2$ pellet is very much less compared to the granular pellet. The dense microstructure in the sintered $MgB_2$ sample reduces the number density of JJs present in the sample and also makes the existing JJs stronger. The granular pellet is more favorable for weak JJs. So in the granular pellet, the number density of JJ is expected to be more than sintered pellet. Due to the occurrence of a large number of weak Josephson junctions in the granular $MgB_2$, JJ decoupling in the granular $MgB_2$ pellet is more frequent than the sintered pellet. This frequent JJ decoupling is the source of the ac loss in polycrystalline samples; so the granular $MgB_2$ pellet shows more ac loss than the sintered $MgB_2$ pellet. Increasing frequency makes the total number of JJ decoupling per second more [6], so the ac loss due to the JJ decoupling increases linearly with the applied frequency.

To summarize, a close correlation between the ac loss and the microstructures of the superconducting samples is observed. The temperature dependent ac loss measurements in the superconductor $MgB_2$ show a well-defined peak below the critical temperature. We argue that this peak in $MgB_2$ is related to the two-gap nature of its superconducting state and Josephson junction decoupling. AC loss in granular $MgB_2$ sample reduces with increasing magnetic field at certain temperatures below $T_c$. Sintering at $750^0$ for 5 hours, increasing pressure or decreasing grain size reduces the ac loss in granular $MgB_2$ sample in the rf range. Increasing the rf amplitude above a certain critical value can make the ac loss in superconducting state more than its normal state.

**Acknowledgements:**

This work is supported by the Department of Science and Technology, University Grants Commission and the Council of Scientific and Industrial Research, Government of India.

Address for Correspondence: e-mail: subhasis@physics.iisc.ernet.in

**Figure Captions:**

1. Temperature dependent ac loss of MgB$_2$ granular sample (grain sizes less than 30 μm and pressurized at 10 MPa) at zero magnetic field. Solid curves are the corresponding modified Ambegaokar-Baratoff fits for tunnel junction suggested by Xu et al [14] and Clem [15]. The inset shows the ac loss of YBCO granular sample and the solid curve is the corresponding Ambegaokar-Baratoff fit [6].

2. Magnetic field dependent ac loss of MgB$_2$ granular sample (grain sizes less than 30 μm and pressurized at 10 MPa) at different temperatures. The magnetic field is swept in between –150 to 150 Gauss.

3. Temperature dependent ac loss of MgB$_2$ granular sample (grain sizes less than 30 μm and pressurized at 10 MPa) for three different amplitudes (700, 1000, and 1300mV) of applied rf voltage at zero magnetic field.

4. Temperature dependent ac loss of MgB$_2$ granular samples (pressurized at 10 MPa) for three different grain sizes at zero magnetic field: (a) less than 30 μm; (b) less than 20 μm; (c) less than 10 μm. Inset shows the temperature dependent ac loss of MgB$_2$ granular samples (grain sizes less than 30 μm) for three different pressures (30 MPa, 20 MPa, and 10 MPa) at zero magnetic field.

5. Temperature dependent ac loss of both the granular MgB$_2$ pellet and the sintered MgB$_2$ pellet. Inset shows the frequency ($f$) dependence of the ac losses for the two samples at $T = 10$ K and 15 MHz $> f >$ 5 MHz. Solid curves are the linear fits with different slopes.

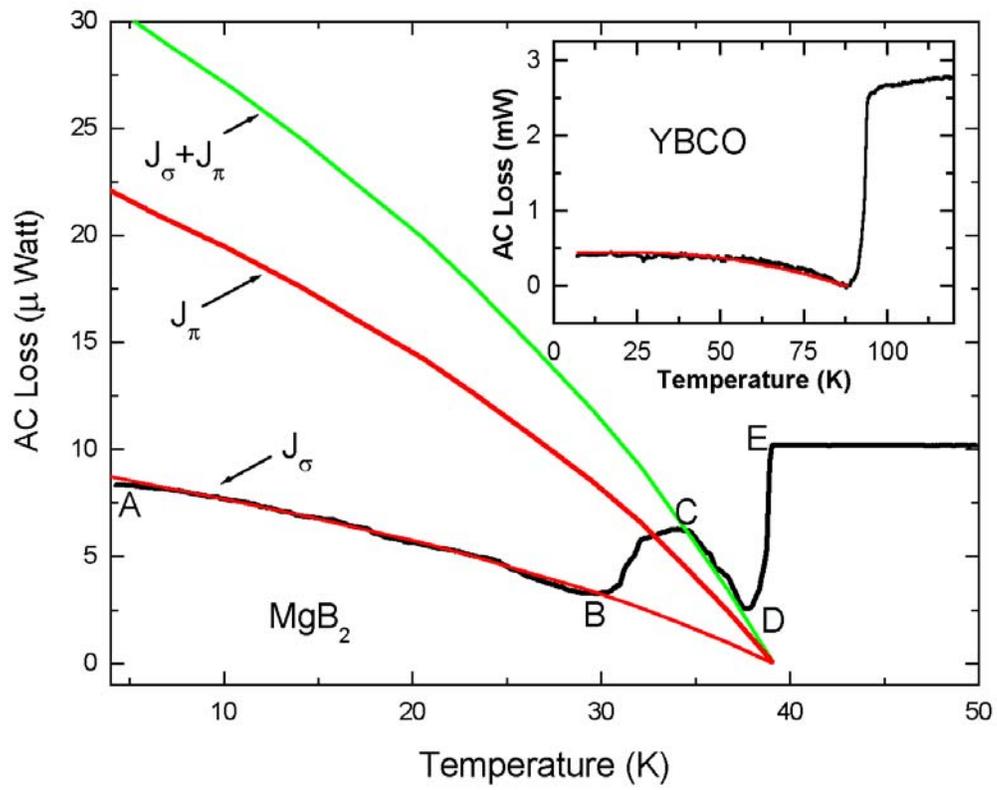

**FIG. 1.**

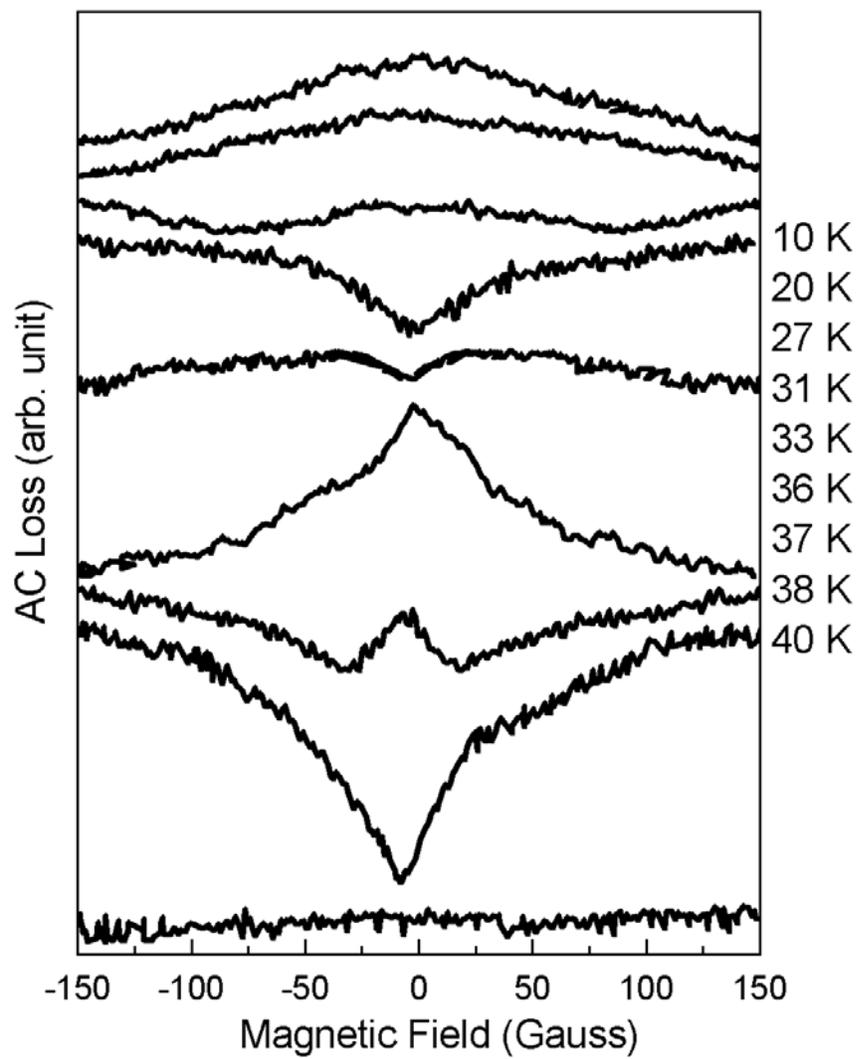

**FIG. 2.**

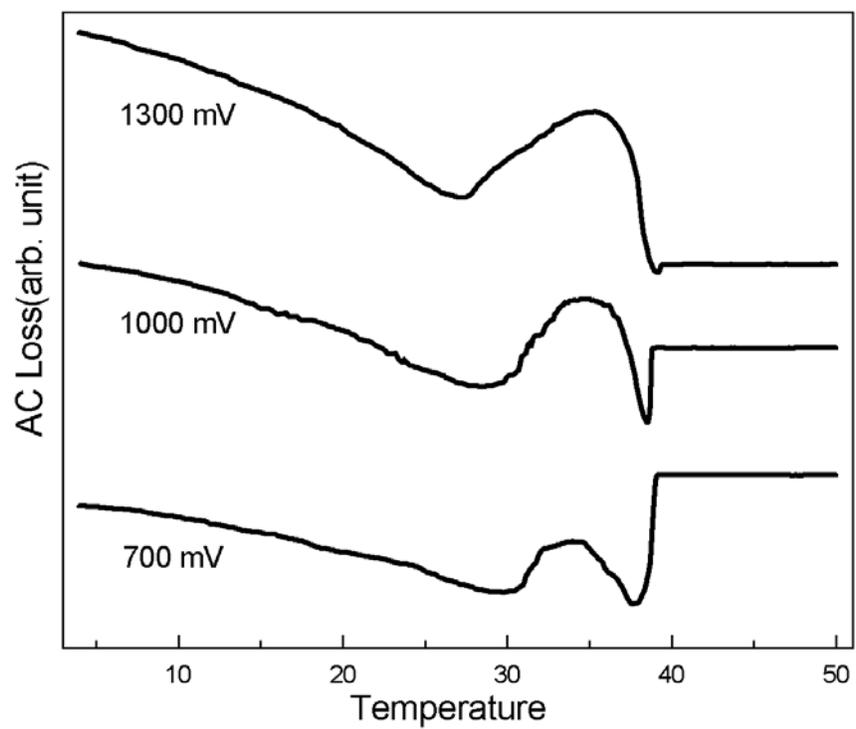

**FIG. 3.**

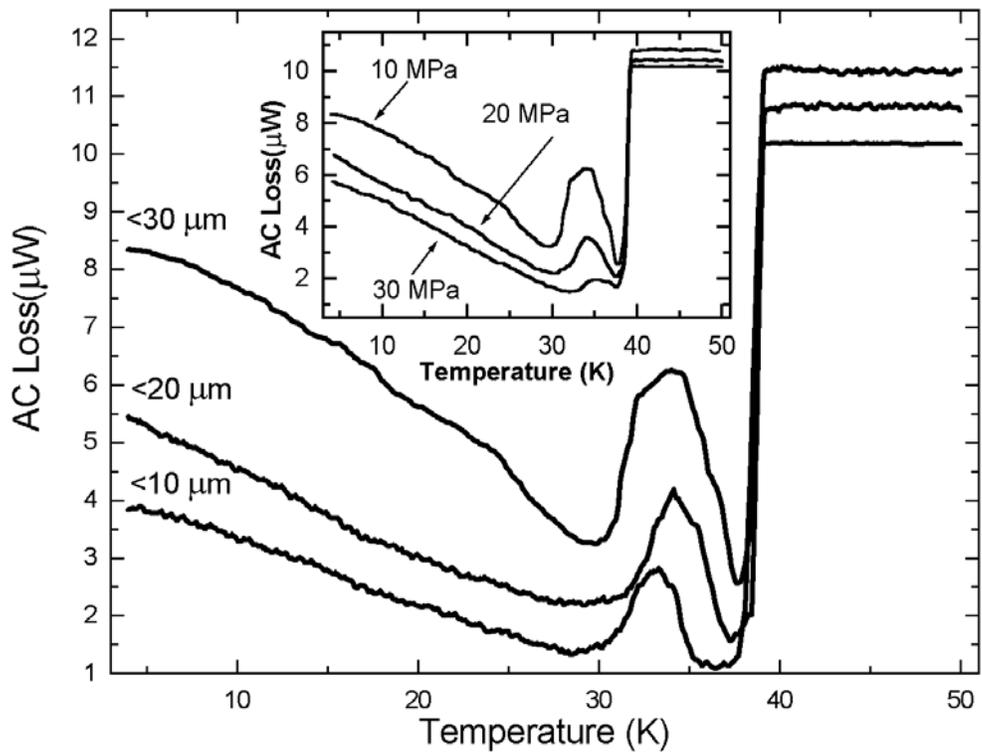

**FIG. 4.**

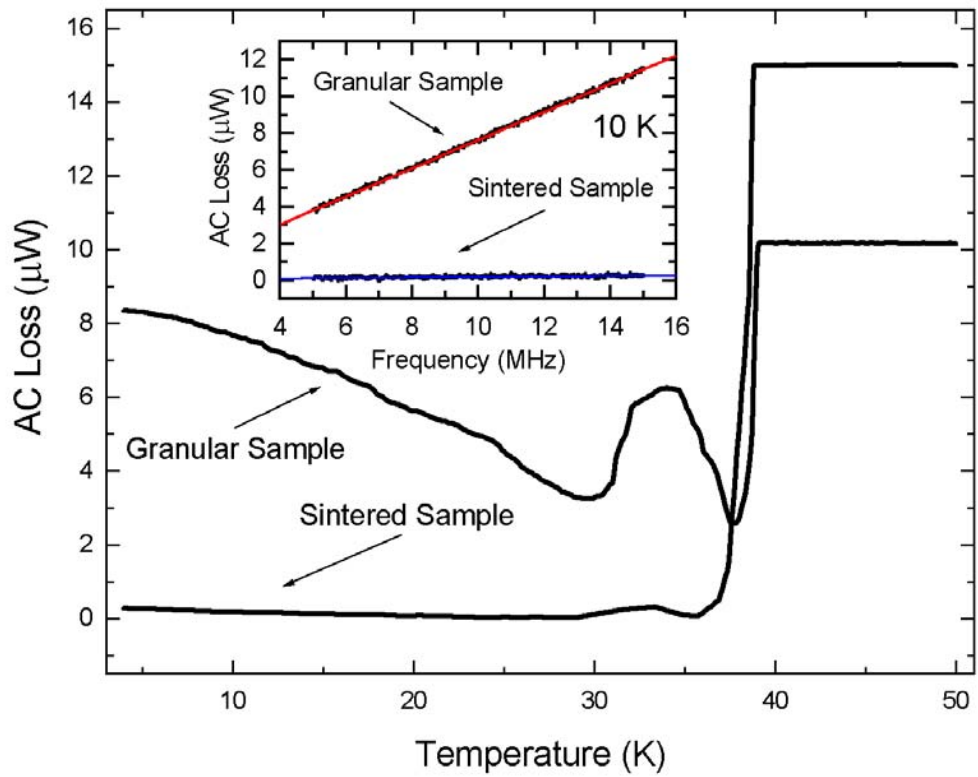

**FIG. 5.**